\documentclass[conference]{IEEEtran}
\IEEEoverridecommandlockouts
\usepackage{cite}
\usepackage{amsmath,amssymb,amsfonts}
\usepackage{algorithmic}
\usepackage{graphicx}
\usepackage{textcomp}
\usepackage{xcolor}
\def\BibTeX{{\rm B\kern-.05em{\sc i\kern-.025em b}\kern-.08em
    T\kern-.1667em\lower.7ex\hbox{E}\kern-.125emX}}

\usepackage{hyperref}
\usepackage{booktabs}
\usepackage{makecell}
\usepackage{subcaption}
\usepackage{minted} 
\usepackage{subcaption}

\usepackage[dvipsnames,table]{xcolor}
\usepackage{textcomp}
\usepackage{multirow}
\usepackage{multicol}
\usepackage{amsmath}
\usepackage{dblfloatfix}
\usepackage{tabularx}
\usepackage[most]{tcolorbox}

\usepackage{xspace}

\def \ie{{\em i.e., }}
\def \eg{{\em e.g., }}

\def \good[#1]{\cellcolor{green!20}{#1}}
\def \limited[#1]{\cellcolor{orange!20}{#1}}
\def \bad[#1]{\cellcolor{red!20}{#1}}

\iffalse
\usepackage[colorinlistoftodos]{todonotes}
\else
\usepackage[disable]{todonotes}
\fi

\newcommand{\michael}[2][inline]{\todo[color=orange!40, #1]{mg: #2}}

\begin{document}

\title{AccelForge: Comprehensive Modeling and Co-Design Framework for AI Accelerators \\ 
}

\author{\IEEEauthorblockN{Tanner Andrulis*}
\IEEEauthorblockA{MIT\\
Cambridge, United States \\
andrulis@mit.edu}
\and
\IEEEauthorblockN{Michael Gilbert*}
\IEEEauthorblockA{MIT\\
Cambridge, United States \\
gilbertm@mit.edu}
\and
\IEEEauthorblockN{Vivienne Sze}
\IEEEauthorblockA{MIT\\
Cambridge, United States \\
sze@mit.edu}
\and
\IEEEauthorblockN{Joel S. Emer}
\IEEEauthorblockA{MIT, NVIDIA\\
Cambridge, United States \\
emer@csail.mit.edu}
}

\maketitle

\begin{abstract}
Tensor algebra workloads, of which deep neural networks are prominent examples, are energy-intensive workloads in modern datacenter and edge deployments, making accelerators necessary to achieve energy efficiency and high throughput. To quickly evaluate and iterate on accelerator designs, we need an accelerator modeling framework that captures salient attributes of devices, circuits, architectures, workloads, as well as optimizing the mapping of the workload onto the hardware.

In this paper, we introduce AccelForge, which improves upon existing accelerator modeling frameworks in capabilities, speed, and ease-of-use. AccelForge unifies and multiple works into one framework, and it includes (1) composable user-defined and user-modifiable models of devices, circuits, and architectures, (2) fast mappers that enable accurate evaluation in orders of magnitude less (computer and human) time, and (3) easy-to-use and easy-to-extend, yet still high performance, Python implementations of both the model and mapper to enable rapid research and extension to novel optimizations.
\end{abstract}


\section{Introduction}
\def\thefootnote{*}\footnotetext{These authors contributed equally to this work}\def\thefootnote{\arabic{footnote}}

Tensor algebra workloads, of which deep neural networks are prominent examples, are energy-intensive workloads in modern datacenters and edge deployments, making it critical to design energy-efficient, high-throughput hardware to run them. Simultaneously, Moore’s law is slowing down and Dennard scaling is over, so transistor scaling can no longer significantly contribute to the energy and throughput improvements we need.

Instead, an important source of efficiency gains is designing accelerators that leverage optimizations across multiple aspects of the design~\cite{room_at_the_top}. First, there are two aspects of the hardware: \emph{components} (devices and circuits) perform data movement, computations, and other actions; and \emph{architectures} organize components into a larger system (\eg how many processing elements, how many memory levels). The next aspect is the \emph{workload}, including the data and the operations (\eg multiply-accumulate) on those data. The last aspect, \emph{mapping}, programs the workload on the hardware (both temporally and spatially)~\cite{cimloop}.



\begin{figure}
    \centering
    \includegraphics[width=\linewidth]{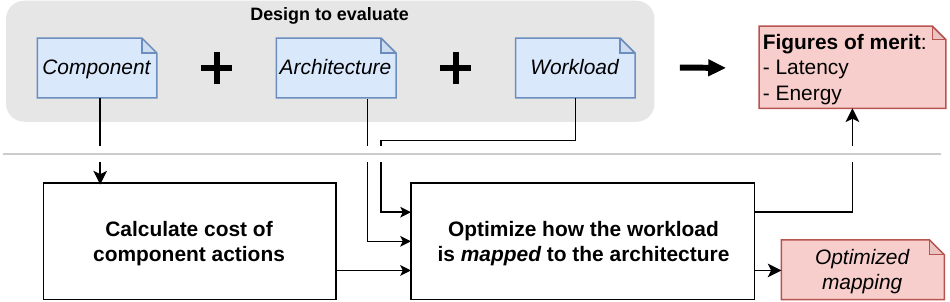}
    \caption{The inputs (blue) and outputs (red) of an evaluation framework, and the processes necessary for evaluation (white).}
    \label{fig:eval_process}
\end{figure}

\michael{Update this paragraph to give the right mapping impression (the full one).}
Such exploration requires a framework to evaluate hardware designs running a workload (\eg estimate the energy of an accelerator running an LLM). Fig.~\ref{fig:eval_process} shows a common evaluation pipeline~\cite{timeloop,cimloop}. The framework takes in the components, architecture, and workload to evaluate. Then, evaluation includes two steps. First, we model component action costs (\eg a buffer read costs 1pJ/bit). Second, informed by action costs, we select a mapping that maps the workload onto the architecture~\cite{timeloop}. Mapping optimization involves search over a space of mappings (\ie \emph{mapspace}), returning the best-found mapping. From this mapping, we can derive action counts (\eg number of buffer reads) and overall costs (\eg total buffer energy) incurred by each component when the hardware executes the mapping.

Based on years of work using and extending~\cite{cimloop,looptree} popular architecture modeling frameworks Timeloop~\cite{timeloop} and Accelergy~\cite{accelergy}, we have distilled three specific goals for an evaluation framework.






\textbf{Goal 1: Supports a sufficiently comprehensive space in all aspects.}
We would like the evaluation framework to explore a comprehensive space of choices in all aspects. ``Comprehensive" here means including the space of choices to explore state-of-the-art optimizations; for example, prior work reduced energy and latency by combining analog and digital components into heterogeneous architectures~\cite{raella,h2_llm}, or by including fused~\cite{convfusion,fusedcnn,optimus,looptree,looptree_thesis} uneven~\cite{zigzag}, and hybrid mappings. Moreover, we aim to be comprehensive in all aspects for at least two reasons. First, co-optimization across several aspects can significantly reduce energy and latency~\cite{raella,h2_llm}. Second, comprehensive evaluations are often necessary. For example, accelerators are used for a wide variety of current and anticipated workloads, and latency and energy can vary significantly between workloads~\cite{timeloop,looptree,zigzag}. Comprehensive evaluation requires an evaluation framework that supports a multitude of workloads. 


\textbf{Goal 2: Fast and optimal mapper.}
We have included in Goal 1 a comprehensive space of mappings. In addition, an evaluation framework must also explore the space of mappings to find the optimal mapping. Prior work has shown that finding suboptimal mappings leads to higher latency and energy for a given design~\cite{zigzag,looptree,tcm}. Moreover, finding suboptimal mappings also misleads designers (\eg incorrectly bias towards overprovisioned hardware resources that could have been used more efficiently by better mappings~\cite{tcm}).


\textbf{Goal 3: Easy to use and extend.}
Each design aspect has a wide space of choices and many configurable parameters. An easy-to-use framework must be modular such that changing one parameter (\eg bitwidth of a tensor element) should require updating one element of the input specification. Moreover, the specification should be intuitive (\eg it should be immediately clear what should be changed to increase memory capacity), and any error in the specification should be identified and an informative error message provided to the user. Finally, because it is impossible to predict future design spaces, the framework must be easy to extend to include more mappings, faster mappers, or to model additional hardware features.

Unfortunately, prior works, as shown in Table~\ref{tab:comparison}, do not meet most of these goals. Although prior work often expands some aspects of the design space, none expand in all of them. Moreover, a fast and optimal mapper has not been devised for such a comprehensive design space. Finally, many prior frameworks are implemented in C++ to achieve the necessary performance to explore the mapspace (albeit still slowly and without finding optimal mappings), which complicates extension.

\def \good[#1]{\cellcolor{green!20}{#1}}
\def \limited[#1]{\cellcolor{orange!20}{#1}}
\def \bad[#1]{\cellcolor{red!20}{#1}}
\newcommand{\ml}[1]{\begin{tabular}{c}#1\end{tabular}}

\begin{table*}[]
    \centering
    \begin{tabular}{c c c c c c c c}

    \multirow{2}{*}{Framework} & \multicolumn{4}{c}{\textbf{Goal 1: Comprehensive design space}} & \multicolumn{2}{c}{\makecell{\textbf{Goal 2: Fast+optimal mapper}}} &  \multirow{2}{*}{\textbf{\ml{Goal 3: Easy to\\use and extend}}} \\
    \cmidrule(lr){2-5} \cmidrule(lr){6-7}
                  & Components & Architecture & Workload & Mapping & Fast Mapper & Optimal Mapper \\
    \midrule
    TileFlow~\cite{tileflow}   & \limited[+Configurable] & \bad[] & \good[Multiple ops] & \limited[+Fusion] & \bad[] & \bad[] & \bad[No] \\
    ZigZag~\cite{zigzag}       & \bad[Baseline] & \bad[] & \bad[One op at a time]  & \limited[+Uneven]      & \bad[] & \bad[] & \good[Yes] \\
    Timeloop~\cite{timeloop}   & \limited[+Configurable] & \bad[] & \bad[One op at a time]  & \bad[Baseline]       & \bad[] & \bad[] & \bad[No] \\
    \makecell{CiMLoop~\cite{cimloop}}     & \limited[\makecell{+Configurable +Leakage}] & \bad[]\multirow{-4}{*}{\makecell{Homogeneous}} & \bad[One op at a time] & \bad[Baseline]  & \bad[]\multirow{-4}{*}{No} & \bad[]\multirow{-4}{*}{No}  & \bad[No] \\
    \midrule
    AccelForge & \good[]\makecell{+Configurable +Leakage\\+Timing} & \good[]\makecell{Heterogeneous} & \good[Multiple ops] & \good[]\makecell{+Fusion\\+Uneven\\+Hybrid} & \good[Yes] & \good[Yes] & \good[Yes] \\
    \bottomrule
    \end{tabular}
    \caption{Comparison of accelerator design space exploration frameworks in accomplishing the three goals presented. For Goal 1, support for some key design features is compared. Only AccelForge accomplishes the three goals of supporting a comprehensive design space, fast and accurate mapper, and being easy to extend. Terminology in Goal 1---\textbf{Configurable}: components can be changed, \textbf{Leakage}: modeling of leakage power, \textbf{Timing}: modeling of component latency and throughput, \textbf{Homogeneous}: composed of the same unit types, \textbf{Heterogeneous}: composed of different unit types, \textbf{Single op}: workloads with one tensor operation, \textbf{Multiple ops}: workload with a graph of multiple tensor operations and inter-operation data dependencies, \textbf{Fusion}: reusing data between tensor operations on-chip, \textbf{Uneven}: per-tensor choice of tile shapes, \textbf{Hybrid}: per-tensor choice of tile shapes between tensor operations.}
    \label{tab:comparison}
\end{table*}

To address this challenge, this paper describes \emph{AccelForge},\footnote{Pronounced ack-sell-forge. \textit{Accel}, as a subset of \textit{Accelerator}, plus \textit{Forge}.} a modeling and DSE framework that incorporates and extends multiple prior works~\cite{cimloop,looptree,looptree_thesis,ffm,tcm} to achieve these goals. The following Section~\ref{sec:design} explains how AccelForge is designed achieve these goals.\footnote{AccelForge is available at \href{https://github.com/Accelergy-Project/accelforge}{https://github.com/Accelergy-Project/accelforge}.} Lastly, Section~\ref{sec:citation} provides guidelines for citing AccelForge as a whole and for citing specific parts of AccelForge. 

\section{Citing AccelForge}\label{sec:citation}
If you would like to cite AccelForge (in general or a specific part), we provide guidelines here.

If you would like to cite the AccelForge framework in general, you can cite this paper you are reading (BibTeX below). TODO: update this
\begin{lstlisting}[basicstyle=\ttfamily\tiny,breaklines=true,columns=fullflexible]
@INPROCEEDINGS{cimloop,
author={Andrulis, Tanner and Emer, Joel S. and Sze, Vivienne},
booktitle={2024 IEEE International Symposium on Performance Analysis of Systems and Software (ISPASS)},
title={CiMLoop: A Flexible, Accurate, and Fast Compute-In-Memory Modeling Tool},
year={2024},
volume={},
number={},
pages={10-23},
doi={10.1109/ISPASS61541.2024.00012}}
\end{lstlisting}

If you would like to cite a more specific part of AccelForge, we list the various components below.
\begin{itemize}
    \item Workload specification.
    \item Architecture specification.
\begin{lstlisting}[basicstyle=\ttfamily\tiny,breaklines=true,columns=fullflexible]
@INPROCEEDINGS{cimloop,
author={Andrulis, Tanner and Emer, Joel S. and Sze, Vivienne},
booktitle={2024 IEEE International Symposium on Performance Analysis of Systems and Software (ISPASS)},
title={CiMLoop: A Flexible, Accurate, and Fast Compute-In-Memory Modeling Tool},
year={2024},
volume={},
number={},
pages={10-23},
doi={10.1109/ISPASS61541.2024.00012}}
\end{lstlisting}
    \item Component specification and included component cost models.
\begin{lstlisting}[basicstyle=\ttfamily\tiny,breaklines=true,columns=fullflexible]
@software{hwcomponents,
  author={Andrulis, Tanner},
  title={HWComponents},
  url={https://github.com/Accelergy-Project/hwcomponents},
  license={MIT},
}
@INPROCEEDINGS{cimloop,
author={Andrulis, Tanner and Emer, Joel S. and Sze, Vivienne},
booktitle={2024 IEEE International Symposium on Performance Analysis of Systems and Software (ISPASS)},
title={CiMLoop: A Flexible, Accurate, and Fast Compute-In-Memory Modeling Tool},
year={2024},
volume={},
number={},
pages={10-23},
doi={10.1109/ISPASS61541.2024.00012}}
\end{lstlisting}

    \item Mapping specification (LoopTree~\cite{looptree}).
\begin{lstlisting}[basicstyle=\ttfamily\tiny,breaklines=true,columns=fullflexible]
@INPROCEEDINGS{looptree,
author={Gilbert, Michael and Wu, Yannan Nellie and Parashar, Angshuman and Sze, Vivienne and Emer, Joel S.},
booktitle={2023 IEEE International Symposium on Performance Analysis of Systems and Software (ISPASS)},
title={LoopTree: Enabling Exploration of Fused-layer Dataflow Accelerators},
year={2023},
volume={},
number={},
pages={316-318},
doi={10.1109/ISPASS57527.2023.00038}}
\end{lstlisting}

    \item Intra-Einsum mapper (TCM~\cite{tcm}).
\begin{lstlisting}[basicstyle=\ttfamily\tiny,breaklines=true,columns=fullflexible]
@misc{turbo_charged,
title={The Turbo-Charged Mapper: Fast and Optimal Mapping for Accelerator Modeling and Evaluation},
author={Michael Gilbert and Tanner Andrulis and Vivienne Sze and Joel S. Emer},
year={2026},
eprint={2602.15172},
archivePrefix={arXiv},
primaryClass={cs.AR},
url={https://arxiv.org/abs/2602.15172}}
\end{lstlisting}
    \item Inter-Einsum mapper (FFM~\cite{ffm}).
\begin{lstlisting}[basicstyle=\ttfamily\tiny,breaklines=true,columns=fullflexible]
@misc{fast_fusiest,
title={Fast and Fusiest: An Optimal Fusion-Aware Mapper for Accelerator Modeling and Evaluation},
author={Tanner Andrulis and Michael Gilbert and Vivienne Sze and Joel S. Emer},
year={2026},
eprint={2602.15166},
archivePrefix={arXiv},
primaryClass={cs.AR},
url={https://arxiv.org/abs/2602.15166}}
\end{lstlisting}
\end{itemize}

\bibliographystyle{IEEEtran}
\bibliography{refs.bib}

\end{document}